\documentclass{iaus}
\usepackage{graphicx}

\title[Outer protoplanetary disks] 
{Chemistry and line emission of outer protoplanetary disks}

\author[I. Kamp et al.]   
{Inga Kamp$^1$,
 Cornelis P. Dullemond$^2$, Michiel Hogerheijde$^3$, Jesus Emilio Enriquez$^4$}

\affiliation{$^1$Space Telescope Division of ESA, STScI, 3700 San Martin 
Drive, Baltimore, MD 21218, USA \break email: kamp@stsci.edu\\[\affilskip]
$^2$Max-Planck Institute for Astronomy, K\"{o}nigstuhl 17, D-69117 
Heidelberg, Germany\\[\affilskip]
$^3$Leiden Observatory, PO Box 9513, NL-3200 RA Leiden, The Netherlands\\[\affilskip]
$^4$University of Texas at El Paso, PSCI 210, 500 W. University Ave., El Paso, TX 79968, USA}

\pubyear{2005}
\volume{231}  
\pagerange{}
\date{?? and in revised form ??}
\jname{Proceedings Astrochemistry: Recent Successes and Current Challenges, IAU Symposium}
\editors{???, eds.}
\begin{document}

\maketitle

\begin{abstract}
The structure and chemistry of protoplanetary disks depends strongly on the 
nature of the central star around which it has formed. The dust temperature 
is mainly set by the stellar luminosity, while the chemistry of the upper 
disk layers depends on the amount of intercepted UV and X-ray flux. 
We will study the differences in chemistry, thermal sturcture and line 
emission around Herbig Ae/Be, T Tauri stars and low mass M dwarfs. 
Predictions will be made for future observations with SOFIA and Herschel. 
\end{abstract}

\section{Introduction}\label{sec:intro}

The physical and chemical conditions in protoplanetary disks set the boundary conditions for planet formation. Although the dust component of these disks has been studied in considerable detail (grain composition, size distribution and mineralogy), very little is known about the gas phase. Over the past decade various disk models have been developed to study the chemical composition of protoplanetary disks itself as well as the impact that it has on the physical structure of the disk (\cite[Aikawa et al. 2002]{aikawa:02}, \cite[van Zadelhoff et al. 2003]{vanzadelhoff:03}, \cite[Kamp \& Dullemond 2004]{kamp:04a}, \cite[Nomura \& Millar 2005]{nomura:05}, \cite[Ceccarelli \& Dominik 2005]{ceccarelli:05}).

Current submm observations are capable of studying the chemical composition of a few nearby young disks; radio interferometry allows in some cases to resolve the spatial distribution of molecules and the rotation pattern of the disk. We use these observations to understand the chemistry in the outer regions of protoplanetary disks. However, these observations are also crucial for determining the mass of these disks. So far, the dust has been used as an indirect tracer for the total disk mass. Recent publications (\cite[Shuping et al. 2003]{shuping:03}, \cite[Brittain et al. 2005]{brittain:05}, \cite[Dullemond \& Dominik 2005]{dullemond:05}) have shown that dust grain growth affects these disks already in very early stages; thus deriving the disk mass from the thermal emission of the small dust grains seems inadequate. A more direct approach is the detection of suitable gas tracers and the proper conversion of emission fluxes into total disk mass. The caveat here is the need of an adequate physical {\em and} chemical model of the disk.

\section{Disk models}\label{sec:models}

The modeling approach is described in \cite[Kamp \& Dullemond (2004)]{kamp:04a} for the
optically thick models and \cite[Kamp \& Bertoldi (2000)]{kamp:00} and \cite[Kamp \& van Zadelhoff (2001)]{kamp:01} for the optically thin models.

The standard disk parameters for all models are a surface density profile $\Sigma(r) = \Sigma_0 (r/{\rm AU})^{-1}$ with $\Sigma_0 = 50$~g~cm$^{-3}$, an inner radius $R_{\rm in} = 0.1$~AU and an outer radius of $R_{\rm out}=300$~AU. The parameters of the central stars are listed in Table~\ref{tab:stars}. The A dwarf models are computed for a late phases of protoplanetary disks, when the disk starts to become optically thin (see \cite[Kamp \& Bertoldi 2000]{kamp:00} for details). The disk masses for these models are ranging from $1.5\times 10^{-4}$ to $1.5\times 10^{-7}$~M$_\odot$ (50 - 0.05 M$_{\rm Earth}$) and the inner and outer radius are 40 and 500 AU respectively. Such inner hole sizes are frequently observed in later stages of disk evolution and often attributed to dust dynamics and/or planet formation.

\begin{table}[h]
\begin{center}
\caption{Stellar parameters used in the disk modeling}
\vspace*{2mm}
\begin{tabular}{lcccc}
Parameter                &   PMS M star   &  T Tauri star  &  Herbig Ae star  &  A dwarf \\[0mm]
\hline\\[-2mm]
$T_{\rm eff}$ [K]      &   3500              &  4000             &  9500                    &  8200 \\
$R_\ast /R_\odot$  &   0.84                &  2.50              &  2.40                     &   1.70 \\ 
$M_\ast /M_\odot$ &    0.5                 &  0.5                 &  2.50                     &    2.0 \\
$L_\ast /L_\odot$   &   0.1                  & 1.5                  & 13.2                      &  12.0 \\
\end{tabular}
\label{tab:stars}
\end{center}
\end{table}

The dust grains in the optically thick models are $0.1~\mu$m astronomical silicate grains (see \cite[Dullemond et al. (2002)]{dullemond:02} for more details). The later phases use $3 \mu$m
black body grains (see  \cite[Kamp \&Bertoldi (2000)]{kamp:00} for more details).

\section{Stellar models}\label{sec:stellar}

\begin{figure}[t]
\begin{center}
\includegraphics[scale=0.7]{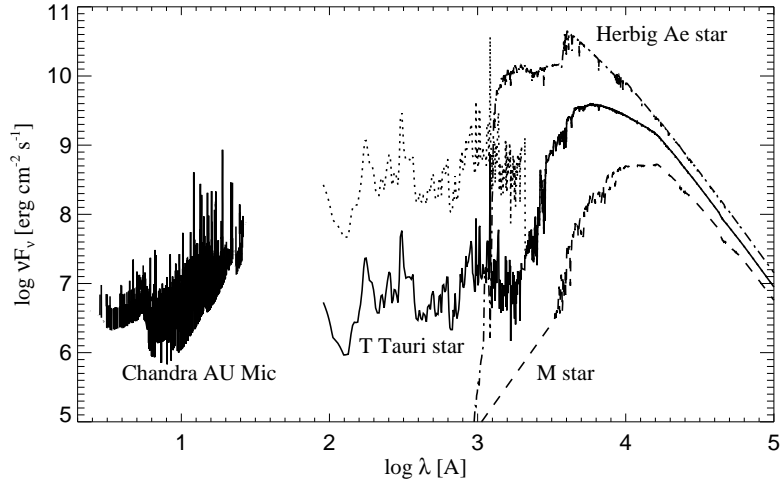}
\end{center}
\caption{Full spectrum of all template stars: T Tauri star with chromosphere (solid line), T Tauri star with an accretion rate of $10^{-9}$~M$_\odot$/yr (scaled chromosphere; dotted line), M star (dashed line) and Herbig Ae star (dash-dotted line); the Chandra spectrum of AU Mic consists only of emission lines.}
\label{fig:stellarUV}
\end{figure}

The dust temperature is mainly set by the total luminosity of the star; hence the optically thick
disk models use a blackbody radiation field with the corresponding effective temperature of the star. On the other hand, the chemistry and heating of the gas is mainly driven by UV photons and X-rays. Line blanketing in stellar atmospheres and chromospheric/accretion activity can cause the UV fluxes to differ by several orders of magnitude from those of a blackbody with the respective effective temperature.
Thus, Kurucz stellar atmosphere models have been used throughout this study together with scaled UV fluxes (according to the age/activity level of the star) and observed X-ray fluxes. For the chromospheric UV excess, we used solar observations scaled with the inverse age $t^{-1}$ (see \cite[Kamp \& Sammar (2004)]{kamp:04b} for details). At an age of 2 Myr, the level of UV irradiation produced by chromospheric activity would roughly correspond to that generated by an accretion rate of
\begin{eqnarray}
\dot{M} & = & 8\, \pi R^2 / v_s^2 {\it F} f \,\,\,{\rm g}~{\rm s^{-1}} \\
              & \approx & 2\times 10^{-11} \,\,\,{\rm M_\odot}~{\rm yr^{-1}}\,\,\, \nonumber
\end{eqnarray}
where $R$ is the inner radius of the disk, $v_s$ is the free-fall velocity at the stellar surface, ${\it F}$ is the accretion flux and $f$ is the filling factor, which is assumed to be 1\% (\cite[Calvet \& Gullbring (1998)]{calvet:98}). To mimic the UV irradiation by a T Tauri star with a typical accretion rate of $10^{-9}$~M$_\odot$/yr, the chromospheric fluxes were simply scaled by a factor 50. 
AU Mic was used as a template for the M star UV and X-ray fluxes. This active M dwarf has been monitored for decades and the respective data was extracted from the HST and Chandra archives. In Fig.~\ref{fig:stellarUV}, the spectra of the three stars, Herbig Ae, T Tauri and M star, have been plotted. The UV spectrum in the M star has been approximated by a linear fit. 

Fig~\ref{fig:stellarUV} shows the Chandra fluxes of AU Mic. For our modeling, we have assumed the Chandra data, because it covers a larger wavelength region than the XMM data. All of these measurements are emission lines and there is no continuum detected. The X-ray heating rate is derived as an integral over the emission spectrum given a simple mean X-ray cross section (\cite[Gorti \& Hollenbach 2004]{hollenbach:05})
\begin{equation}
\sigma(E) = 1.2\times 10^{-22} \left(\frac{E}{{\rm 1~keV}}\right)^{-2.594} \,\,\,{\rm cm}^2 \, ,
\end{equation}
where $E$ is the energy in eV. This cross section is derived for elemental abundances measured through the diffuse cloud toward the star $\zeta$~Oph.

We calculated eight protoplanetary disk models to assess the influence of various parameters like the spectral type of the central star, PAHs and X-rays on the physical and chemical structure of the protoplanetary disks. Model 1 is the Herbig Ae star, Model 2a-d are the various T Tauri star models and 3a and b are the M star models. The properties of these disk models are summarized in Table~\ref{tab:model}.

\begin{table}[h]
\begin{center}
\caption{Overview of the massive disk models: 1 -- Herbig Ae star, 2 -- T Tauri star and 3 -- M dwarf}
\vspace*{2mm}
\begin{tabular}{ccccc}
number  &  star            &  PAH heating &      UV flux         & X-rays \\
  \hline\\[-2mm]
1 & Herbig Ae  &     on                & Kurucz              & none \\
 2a & T Tauri       &      on                & chromosphere & none \\
 2b & T Tauri       &      off                & chromosphere & none \\
 2c & T Tauri       &      on                & accretion          & none \\
 2d & T Tauri       &      off                & accretion          & none \\
 3a & M dwarf     &      on                & AU Mic fit          & AU Mic observation \\
 3b & M dwarf     &      on                & AU Mic fit          & AU Mic observation $\times 0.01$ \\
\end{tabular}
\label{tab:model}
\end{center}
\end{table}

\section{Gas Temperatures in protoplanetary disks}\label{sec:gast}

\begin{figure}[t]
\hspace*{-2mm}\includegraphics[scale=0.76]{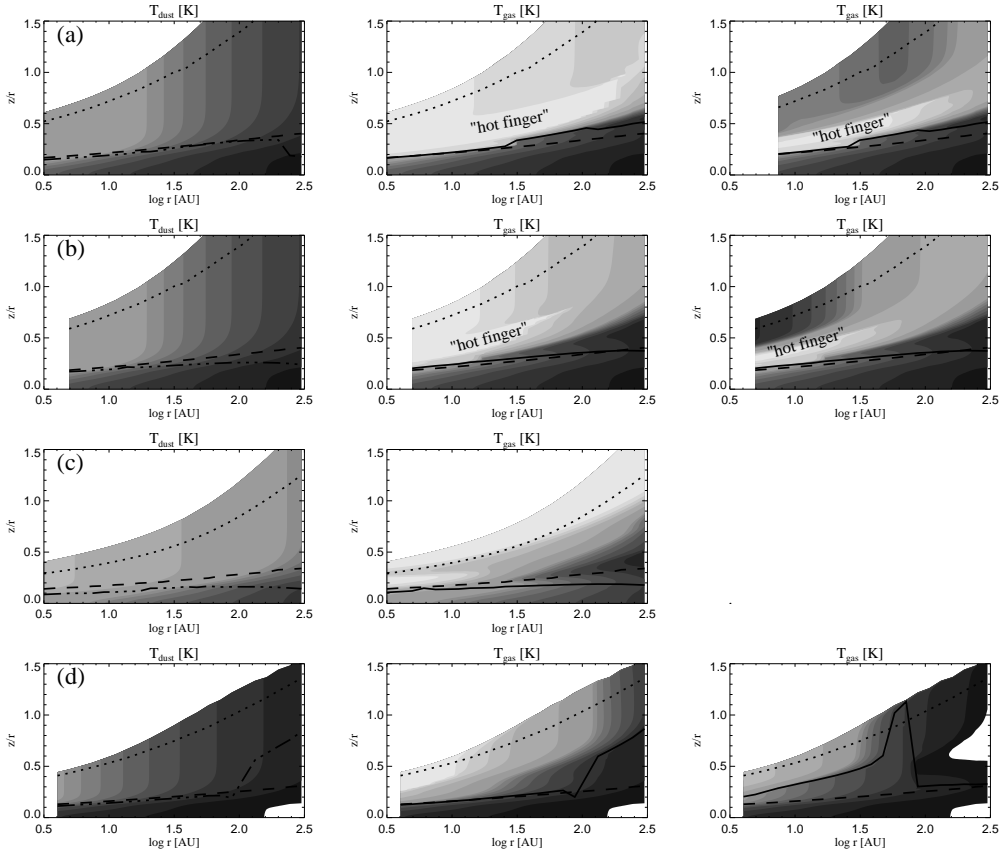}
\caption{Dust and gas temperatures in the massive disk models: (a) model 2c/d: T Tauri star with an accretion rate of $10^{-9}$~M$_\odot$/yr, (b) model 2a/b: T Tauri star with chromosphere, (c) model 1: Herbig Ae star, (d) model 3a/b: Mstar with and without X-rays.
The left column shows the dust temperature, the middle column the gas temperature in the models with PAHs and the right column the gas temperature without PAHs; in the case of the M star model, the right column shows the gas temperature with PAHs as well, but with a hundred times weaker X-ray fluxes.
The solid line indicates the location, below which gas and dust temperatures agree to within 10\%
(middle and right column). If over an interval $\delta z/z \sim 0.3$ both temperatures agree within 5\%,
complete coupling, that is T$_{\rm gas} =$ T$_{\rm dust}$, is adopted below this depth 
(dashed-dotted line in left column). The 10\% agreement occurs at or slightly above the 
$\tau = 1$ surface layer (dashed line). The dotted line marks the 
transition between the surrounding cloud and the disk. The grey levels correspond to the following temperatures: 10, 20, 30, 40, 50, 60, 70, 80, 90, 100, 200, 500, 1000, 2000, and 5000\,K.}
\label{fig:gas}
\end{figure}

Fig.~\ref{fig:gas} shows the dust and gas temperatures for the seven massive disk models. All models have the same surface density and hence the same disk mass of 0.01~M$_\odot$. The only difference between them is the central stellar radiation field. Our modeling makes clear that UV radiation is the most important heating source for the disk. X-ray heating is only important if the UV flux of the central star is low (like in the case of the M star) and/or very small grains like polycyclic aromatic hydrocarbons (PAHs) are absent.

The hot finger in the T Tauri model was first described in \cite[Kamp \& Dullemond (2004)]{kamp:04a} as the hot surface layer just above $\tau=1$ (see indication in Fig.~\ref{fig:gas}). The very low density top layers are again cooler than the hot surface. The feature becomes even more prominent in the model with accretion UV irradiation. In the latter models, the hot finger extends even out to 200~AU. An external UV radiation field from a nearby O or B star will most likely have the same effect. Since such an external radiation field does not depend on the radial distance to the central star, we can speculate that it might turn this hot finger into a closed hot surface layer: a hot skin of several thousand K around the protoplanetary disk.

This hot finger is less pronounced in the Herbig Ae model, even though PAHs are included. This is due to the much lower UV flux from the central star. In the T Tauri model, the remnant material above the disk (parental molecular cloud) had a constant density of $3\times 10^3$~cm$^{-3}$. For the Herbig Ae star, we considered a spherical cocoon with a power law $3\times 10^3 (r/R_{\rm i})^{-1.3}$~cm$^{-3}$. This surrounding material was mainly introduced to avoid boundary effects on the disk surface. Since the density of the cocoon material decreases with distance for the Herbig Ae star, the gas stays hot even though the UV from the central star is diluted. Fig.~\ref{fig:gas} shows that this gas is going to be very hot (5000-10000~K). 

The M star model illustrates the effects of X-ray heating. The UV fluxes of the M star are several orders of magnitude lower than the T Tauri UV fluxes. Fig.~\ref{fig:gas}(d) compares the gas temperature found with the X-ray fluxes taken from AU Mic and with 100 times lower fluxes. Gas temperatures are typically a factor 2-10 lower in the reduced X-ray case.

\section{Chemistry of protoplanetary disks}\label{sec:chem}

The chemical network used in this study is a small one with 48 species and 268 reactions. It is well suited to calculate the abundances of the relevant cooling species and contains the photochemistry important in the upper layers of protoplanetary disks. However, X-ray chemistry is missing so far and hence the results for the M star should be considered with caution. In the following, we briefly discuss the most important features of these chemical calculations.

\begin{figure}[t]
\hspace*{7mm}\includegraphics[scale=0.7]{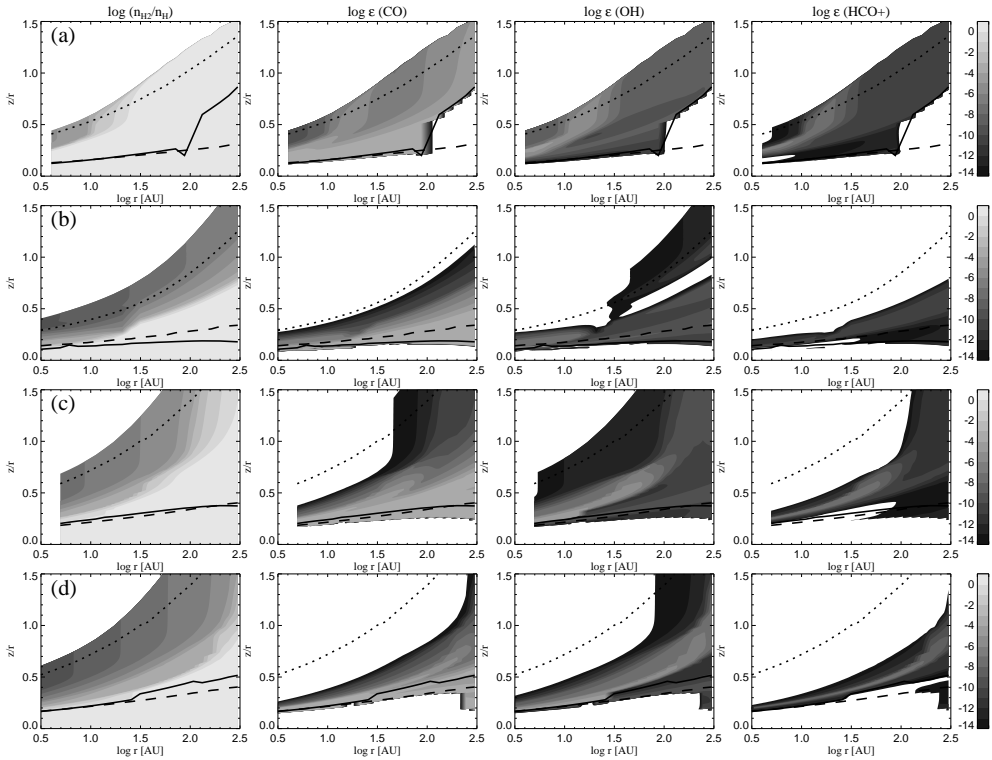}
\vspace*{0.5cm}
\caption{Selected chemical species in four of the seven disk models: (a) M star with X-rays, (b) Herbig Ae star, (c) T Tauri star with chromosphere and PAHs, (d) T Tauri star with an accretion rate of $10^{-9}$~M$_\odot$/yr and PAHs. The first column shows the H$_2$/H fraction, the other columns show the abundances ($\log \epsilon = \log n({\rm species})/n_{\rm tot}$) of CO, OH and HCO$^+$ respectively. }
\label{fig:chem}
\end{figure}

Comparing Fig~\ref{fig:chem}(c) and (d) reveals the impact of the stellar UV irradiation on key molecules like H$_2$ and CO. In the case of pure chromospheric UV flux, H$_2$/H fraction in the outer disks upper layers is 1, meaning that all hydrogen is in molecular form. Also, the C/CO transition occurs at lower optical depth than in the T Tauri accretion model. As discussed above, the hot finger extends to much larger radii in the accretion model and this is also reflected in the OH and HCO$^+$ abundance structure. Both species arise from the fact that H$_2$ is collisionally destroyed by H and O in these layers (\cite[Kamp \& Dullemond 2004]{kamp:04a}). OH is subsequently photodissociated and this process leaves in $\sim 50$~\% of the cases the O-atom in the $^1$D excited state (\cite[St\"{o}rzer \& Hollenbach 1998]{stoerzer:98}). This level gives rise to the [O\,{\sc i}]~6300~\AA\  line. The line can also be thermally excited at temperatures larger than 3000~K. The hot finger presents thus an ideal reservoir of gas for the 6300~\AA\  line emission.

In conditions, where the disk is exposed to an external UV radiation field, like in the Orion nebula, this hot finger might turn into a hot surface of the disk. Actually, observations of the proplyds in Orion with a narrow band filter centered on the 6300~\AA\  line have revealed that this emission comes from the "skin" of the disk (\cite[Bally et al. 1998]{bally:98}). The same line emission has also been observed in Herbig Ae/Be stars by \cite[Acke et al. (2005)]{Acke:05}. These authors are able to explain the emission using an OH abundance of $10^{-6} - 10^{-7}$ and non-thermal emission. Our model, however, suggests that Herbig Ae stars show a hot finger similar to T Tauri stars; thus thermal emission might contribute significantly to the total emission and help to explain the [O\,{\sc i}]~6300~\AA\  line emission with smaller OH abundances and thus abundances closer to the ones derived in our modeling ($10^{-8}-10^{-9}$).
   
\cite[Aikawa et al. (1998)]{aikawa:1998} have shown that X-ray chemistry mainly enhances the presence of ions and thus may affect species that are formed via ion-molecule reactions. However, species like CO seem to be rather unaffected and their abundance is mainly determined by the level of UV irradiation. We abstain from a detailed discussion of the observational consequences arising from the chemistry displayed in Fig.~\ref{fig:chem}, and restrict ourselves instead to a purely theoretical interpretation. It is interesting to note the rather large abundance of molecular hydrogen and CO in this disk model. This is due to the extremely low UV luminosity in the M star model. Even though we used an active star, namely AU Mic, as a template for this model, the total luminosity is much smaller than in T Tauri stars of similar activity level. The maximum of the OH abundance distribution coincides with the end of the hot X-ray heated surface. Apparently, the chemistry there is very similar to what we noted for the hot finger.

\section{Chemistry of more evolved disks}\label{sec:evolved}

As the protoplanetary disks evolve, the dust grows into large grains and protoplanetesimals; the total mass in large grains (micron-sized) is much smaller than the mass in the original dust disk and the entire disk becomes optically thin to continuum radiation. However, H$_2$ and CO photodissociation are not continuum processes and thus significant amounts of these molecules can lead to an efficient shielding of the UV radiation within the disks.
\begin{figure}[ht]
\vspace*{0.0cm}
\hspace*{-2.2cm}\includegraphics[scale=0.62]{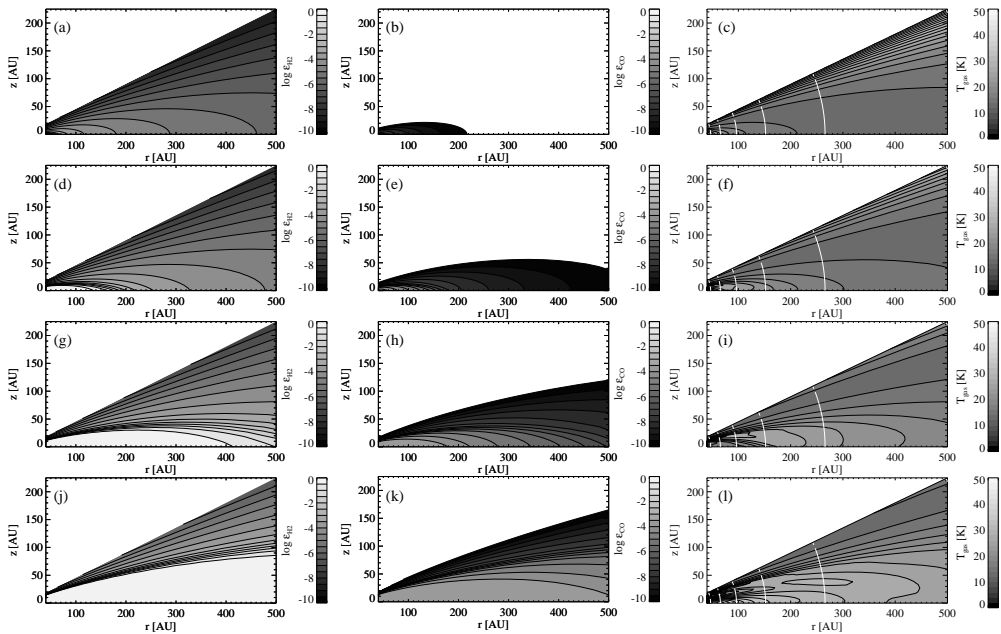}
\vspace*{-1.6cm}
\caption{H$_2$ and CO abundances in the optically thin disk models and gas and dust temperatures: (a-c) $1.5\times 10^{-7}$~M$_\odot$, (d-f) $1.5\times 10^{-6}$~M$_\odot$, (g-i) $1.5\times 10^{-5}$~M$_\odot$, (j-l) $1.5\times 10^{-4}$~M$_\odot$. Contour lines show abundances $\log \epsilon = n_{\rm species}/n_{\rm tot}$ in logarithmic steps. The temperature plots show the gas temperature as filled contours and the dust temperature overplotted with white contour lines (80, 70, 60, 50, 40~K from inside out) }
\label{fig:thin}
\end{figure}

Fig.~\ref{fig:thin} illustrates the abundances of H$_2$ and CO in four disk models spanning a range from $1.5\times 10^{-4}$ to $1.5\times 10^{-7}$~M$_\odot$. The stellar radiation field is that of an A dwarf (8200 K, $\log g = 4.0$ and the interstellar radiation field corresponds to the one in the solar neighbourhood ($\chi = 1$). Even though the disk masses span 4 orders of magnitude, the CO densities in these disks span a much larger range. This is due to strong photodissociation in the lowest disk mass models. As the disk mass increases, shielding by H$_2$ and CO becomes more efficient until all carbon is in the form of CO ($1.5\times 10^{-4}$~M$_\odot$ model). While the CO/H$_2$ conversion factor in the most massive disk model is close to the canonical value of $10^{-4}$ of molecular clouds, it can be as low as $10^{-7}$ in the lowest mass model.

\section{Pushing future observations to the limits}\label{sec:future}

We used the 2D Monte Carlo code from \cite[Hogerheijde \& van der Tak (2000)]{hogerheijde:00} to calculate the line emission from the optically thin disk models. These fluxes are especially interesting in the context of the new upcoming observing facilities like APEX, ALMA, Herschel and SOFIA. We placed the disks at a typical distance of 100~pc and inclined them by $45^{\rm o}$.

\begin{figure}[t]
\includegraphics[scale=0.38]{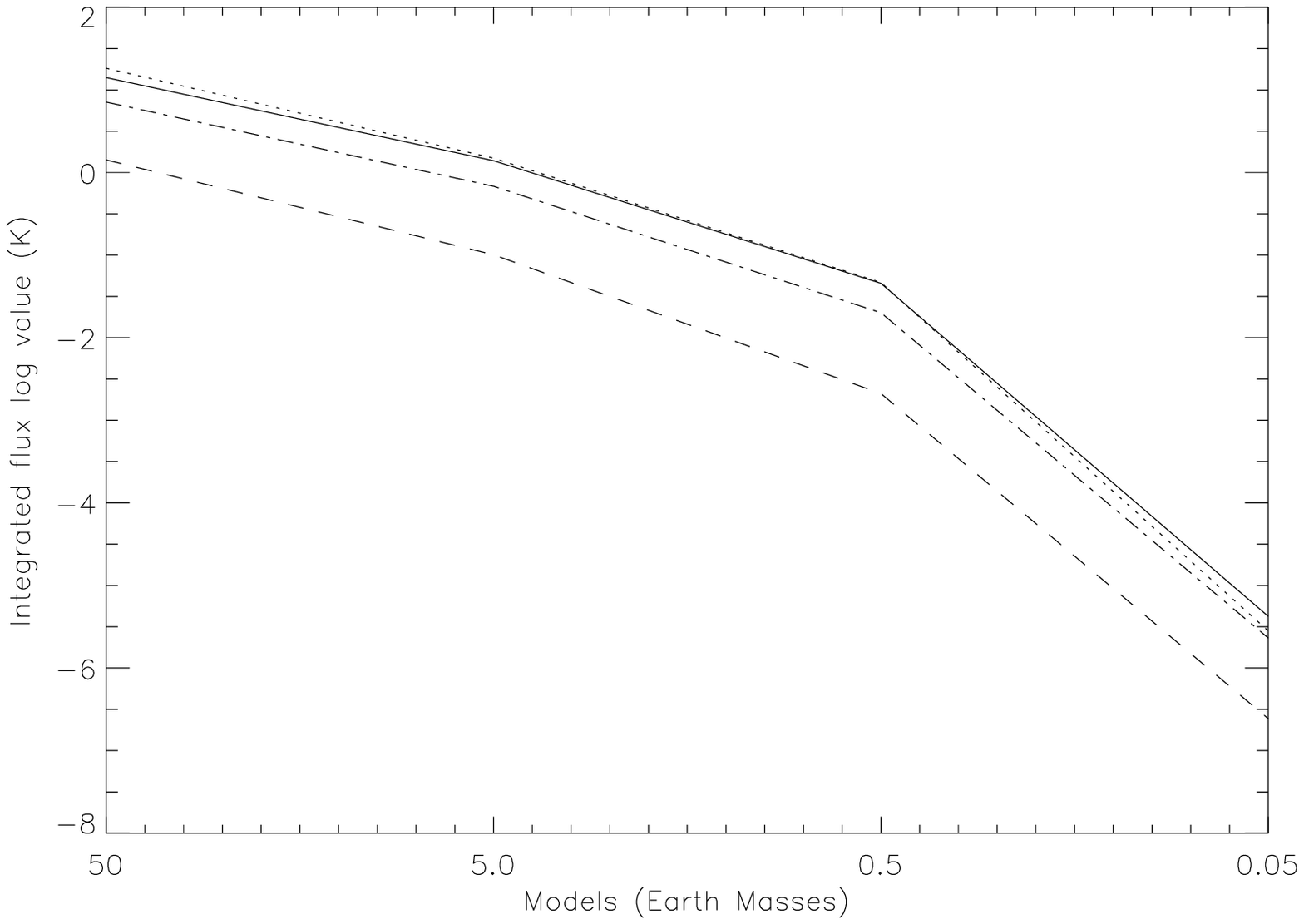}
\includegraphics[scale=0.38]{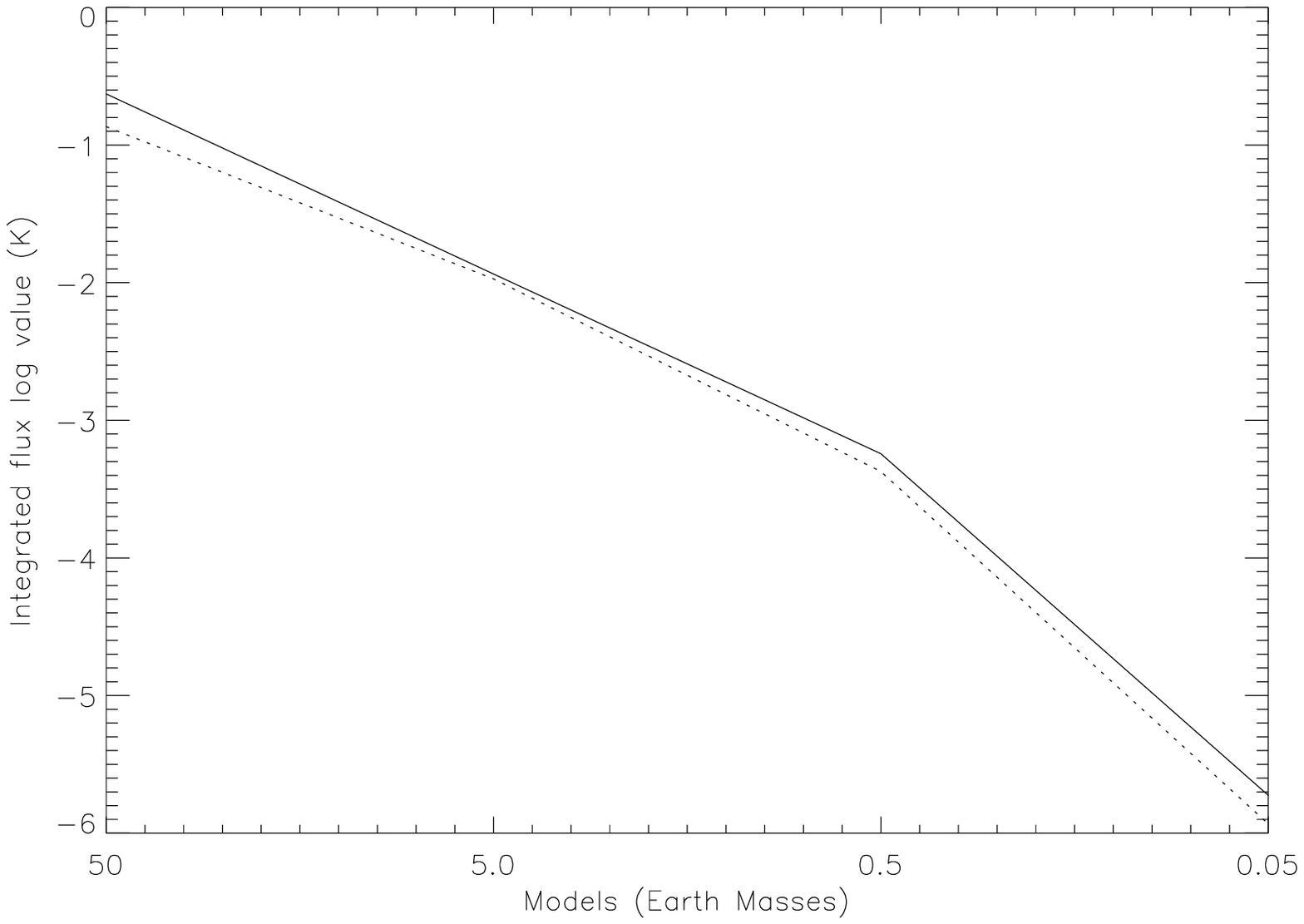}
\vspace*{-0.3cm}
\caption{Integrated CO (left) and O (right) line fluxes from the optically thin disk models: the distance is assumed to be 100~pc and the inclination is $45^{\rm o}$. Shown are the lowest four rotational lines of CO, J=1-0 (dashed), J=2-1 (dash-dotted), J=3-2 (dotted), J=4-3 (solid) and the two fine structure lines of neutral oxygen, 63~$\mu$m (solid) and 145~$\mu$m (dotted)}
\label{fig:COlines}
\end{figure}

For the lowest four CO rotational lines J=1-0, 2-1, 3-2, 4-3, we assumed typical beam widths of 43, 22, 14, and 11" respectively. At a distance of 100~pc, this means that the beam contains the entire disk. Fig~\ref{fig:COlines} shows that the higher CO rotational lines are better to detect, because the corresponding levels are more populated at the typical disk gas temperatures of 30-50~K. This figure illustrates also that the slope of the line fluxes as a function of disk mass is much steeper than 1. As the disk masses become lower than 0.5~M$_{\rm Earth}$, the gas becomes rapidly undetectable in CO even with ALMA.

For the two fine structure lines of neutral oxygen at 63~$\mu$m and 145~$\mu$m, we assumed beam width of 7" (SOFIA/ GREAT) and 9" (Herschel/PACS) respectively. Assuming a typical detection limit of 5 mK in 5 hours of integration time, we can expect to detect disk masses down to a few Earth masses. The [O\,{\sc i}] fine structure lines are thus not as sensitive as the CO lines with ALMA. However, we expect the [C\,{\sc ii}] 158~$\mu$m line to be even stronger than [O\,{\sc i}] for these tenuous disks.

\section{Conclusion}

The interpretation of current near infrared and submm observations needs self-consistent disk models, like the ones illustrated in the previous sections. Disk structure and chemistry together can be used to derive quantities such as disk gas mass and inclination. The power of such disk models as a tool for observations has been recently illustrated in various papers (\cite[Semenov et al. 2005]{semenov:05}, \cite[Jonkheid et al. 2005]{jonkheid:05}, \cite[Nomura \& Millar 2005]{nomura:05}). 

We showed the importance of a proper inclusion of all radiation sources --- stellar UV, X-rays and external irradiation --- for the gas temperature and chemical structure of the disk. The chemistry of the outer protoplanetary disk layers is driven by irradiation; thus it is important to take into account the shape of the radiation field and not only its strength. The wavelength distribution of UV photons in a T Tauri star is very different from that in a mean interstellar radiation field. Instead of using the UMIST photorates that were derived for a Draine field, we conclude that it is necessary to integrate the respective photoionization and dissociation cross sections over the stellar spectrum.

An important step in understanding disk evolution and in setting the boundary conditions for planet formation, is the measurement of the disk gas mass as a function of time and type of central star. Using our disk models, we conclude that future instrumentation like ALMA will allow the detection of transition disks down to 0.5~M$_{\rm Earth}$ of gas.

\section{Outlook}

Disks span a wide range in parameter space --- density, temperature, irradiation --- and the chemical conditions we encounter in them range from extreme PDRs, diffuse clouds to dense clouds and even dark cores. However, these disks become much denser in the midplane than molecular clouds or dark cores and hence three body reactions will play a role there as well. Any model that wants to treat the full chemistry needs to take into account photochemistry, X-ray chemistry, gas-grain chemistry and three body collisions as well as grain surface chemistry. Progress in the future has to be made especially in the field of gas-grain chemistry (a better understanding of desorption processes) and grain surface chemistry. The latter is important in forming very large molecules and more complex organic species, the precursors of life.

In terms of disk modeling, the next generation models should be 2D hydrodynamical models that account for the proper treatment of mixing in the inner disk regions, where diffusion timescales are shorter than or comparable to the chemical timescales. These models would need a realistic energy equation for the gas, because the surface and intermediate layers are not dominated by accretion heating. As outlined above, the full chemistry (gas chemistry, gas-grain and grain surface chemistry) is very expensive and it might be necessary to optimize the chemical networks for the various parameter regimes in the disk.

\begin{acknowledgments}
We would like to thank Jan-Uwe Ness for reducing the AU Mic Chandra data and providing the X-ray fluxes for AU Mic. Floris van der Tak helped with the Monte Carlo radiative transfer code. We also like to thank the STScI for its summer internship program during which part of this work was carried out.
\end{acknowledgments}

\newpage

{\bf Question (Edwin Bergin):} You see evidence for hot fingers with higher temperatures in layers with colder temperatures above. Photoelectric heating is important for the upper cold layers so is the temperature rise due to an increase in photoelectric heating efficiency or lower cooling?\\

{\bf Answer (Inga Kamp):} This temperature inversion occurs in very low density regimes, where the main cooling ([O\,{\sc i}] fine structure lines) is in NLTE and thus scales linearly with density. The photoelectric heating scales as $n^{1.8}$ in this regime. For a vertical slice of the disk, the density slowly increases towards the midplane and thus the heating rises faster than the cooling.\\[1cm]

{\bf Question (Al Glassgold):} There is some question about whether cosmic rays penetrate the entire volume of T Tauri disks due to the fact that they drive winds which sweep away galactic cosmic rays.\\

{\bf Answer (Inga Kamp):} Cosmic rays are mostly important in the midplane layers of the outer disks, where UV irradiation cannot reach. There, they drive the deuterium chemistry and deuterated species are actually observed in protoplanetary disks. However, additional ionization may also come from the stellar X-rays (secondary UV photons) and thus external cosmic rays may be less important to explain the observations. Cosmic rays do not affect the thermal structure of the surface layers in protoplanetary disks.

\end{document}